# AutoSimTTF: A Fully Automatic Pipeline for Electric Field Simulation and Treatment Planning of Tumor Treating Fields

Xu Xie[1], Zhengbo Fan[2], Yue Lan[3], Yun Pan[4], Guangdi Chen[3], Shaomin Zhang[5], Yuxing Wang[1] and Minmin Wang[6]

[1] College of Biomedical Engineering and Instrument Science, Zhejiang University, Hangzhou, China
[2] Polytechnic Institute, Zhejiang University, Hangzhou, China
[3] Department of Public Health, Zhejiang University School of Medicine, Hangzhou, China
[4] College of Information Science and Electronic Engineering, Zhejiang University, Hangzhou, China
[5] Qiushi Academy for Advanced Studies, Zhejiang University, Hangzhou, China
[6] Westlake Institute for Optoelectronics, Westlake University, Hangzhou, China

E-mail: minminwang@zju.edu.cn

## Abstract

*Objective*: Tumor Treating Fields (TTFields) is an emerging approach for cancer therapy that inhibits tumor cell proliferation by applying alternating electric fields (EF) of intermediate frequency and low intensity. The TTFields-induced electric field intensity at the tumor site is closely related to the therapeutic efficacy. Therefore, the EF simulation based on realistic head models have been utilized for the dosage analysis and treatment optimization of TTFields. However, current modeling methods require manual segmentation of tumors and rely on commercial software, which is time-consuming and labor-intensive. *Approach*: We introduce AutoSimTTF, a fully automatic pipeline for simulating and optimizing the EF distribution for TTFields. The main steps of AutoSimTTF utilize open-source toolkits, enabling fully automated processing of individual MRI data for TTFields. Additionally, AutoSimTTF allows for parameter optimization based on individual anatomical information, thereby achieving a more focused and higher EF distribution at the tumor site. *Main results*: Compared to conventional EF calculation processes, deviations in AutoSimTTF are below 20%. The optimal treatment parameters generated by AutoSimTTF produces a higher EF intensity at the tumor site (111.9%) and better focality (19.4%) compared to traditional TTFields settings. *Significance*: AutoSimTTF provides significant reference value and guidance for the clinical application and treatment planning of TTFields.

## 1. Introduction

Tumor Treating Fields (TTFields) is a promising physical approach for tumor treatment. It inhibits the growth of tumor cells by applying alternating electric field (EF) of intermediate frequency (100-500 kHz) and low intensity (1-3 V/cm) through transducers placed on the skin surface. Several basic studies and clinical trials have validated its efficacy and safety [1-3].

The EF intensity generated in the tumor region is closely related to the therapeutic effect of TTFields. The EF distribution in the target region is influenced by factors such as head structure, pathology of tumor, and dielectric properties of tissues. Even with the same transducer montage (location and injected current), the EF distribution produced by TTFields varies significantly between patients, necessitating individual optimization and adjustment of treatment plans. However, due to the technical limitations, it is very difficult to measure the intracranial EF distribution directly. Therefore, numerical simulations are used to predict the distribution of induced EF among the patient.

The individualized magnetic resonance imaging (MRI) images are generally obtained to construct the head model for performing intracranial EF calculations. MRI images are segmented into tissues with different dielectric properties, including scalp, skull, cerebrospinal fluid (CSF), gray matter, and white matter. Each tissue is assigned dielectric properties specific to the frequency, including permittivity and conductivity [4-8]. The transducers are then placed on the head model, and the finite element method is used to calculate the electromagnetic field distribution generated by TTFields.

Creating accurate human head models is a complex and technically intensive task. Current EF modeling workflows are predominantly manual or semi-automatic. Miranda et al. used T1 and proton density-weighted MRI from the Colin27 dataset [9] to create the first head model for TTFields simulations. This segmentation was performed using the BrainSuite [11]. Subsequently, Wenger et al. created a head simulation model using T1, T2, and diffusion MRI of a young healthy female, employing open-source software, SimNIBS v1, for segmentation. The brain cortex was segmented using FreeSurfer [13], and outer tissues were segmented using FSL [14]. However, the model could not segment actual tumors and simulated the presence of tumors by introducing a virtual lesion composed of two concentric spheres [11-12]. Lok et al. segmented MRI of a recurrent GBM patient using commercial software ScanIP to create a realistic head model with tumor [15]. Despite various tools significantly improving the efficiency of head MRI segmentation, time-consuming manual correction is still required post-segmentation, such as removing disconnected voxels, merging unassigned intracranial voxels, and adding transducers on the scalp surface. Timmons et al. developed a semi-automated workflow to simulate TTFields for GBM patients that regions of tumor require manual segmentation by experts. Although the workflow partially automates the procedure, it still requires manual intervention in steps such as tumor segmentation and mesh correction, which typically takes around 24 hours for an individual head model [20].

In addition, current treatment planning of TTFields methods also have certain limitations. Traditional TTFields include two pairs of 3×3 transducer arrays placed anterior-posteriorly and left-rightly on the head. These pairs of arrays alternately apply currents, creating EF stimulations in alternating directions within the cranium. This montage generates a very diffuse EF distribution and cannot provide precise treatment for each tumor site. Therefore, NovoTAL, a commercial software (Novocure, LLC), was designed for individual optimization for clinical treatment. NovoTAL matches patient morphological measurements such as head size, tumor size, and tumor coordinate sites and selects the array layout that can produce the maximum average EF intensity at the tumor sites [21]. The optimization method does not consider the individualized anatomical information and merely focuses on morphological aspects and is limited to selecting from a solution space with two pairs of 3×3 arrays, which is failed to solve the problem of diffuse EF distribution within the cranium.

In summary, there is currently no fully automated pipeline for constructing head model and optimizing treatment parameter for TTFields. The current modeling processes for TTFields require manual correction after automatic segmentation. Tumor regions are often replaced with ideal spherical substitutes or require manual segmentation by experts, making it difficult to ensure consistent segmentation standards. Existing semi-automated workflows for TTFields involve manual correction of mesh models. Treatment planning methods for TTFields need further improved. Therefore, this paper aims to propose an automated pipeline for EF simulation and treatment planning of TTFields

## 2. Methods

### 2.1 Overall pipeline of AutosimTTF

The fully automated pipeline of EF simulation and treatment planning for TTFields is outlined in figure 1. The workflow primarily includes the following steps: acquiring individual MRI images, segmentation of brain tissues and tumor, transducer placement, mesh generation, finite element calculations, and parameter optimization (or specifying transducers).

### 2.1.1 Individual MRI images acquisition

The University of Pennsylvania Glioblastoma Multiforme (UPenn-GBM) dataset [20] was used for processing in this study. This dataset contains multi-modal MRI from 611 glioblastoma patients at the pre-operative baseline time point, including T1, T2, FLAIR, and T1Gd, as well as diffusion tensor imaging (DTI) and dynamic susceptibility contrast (DSC) MRI images for most cases. Facial information has been removed from all patient images. T1, T2, and T2-FLAIR images are registered with T1Gd images, with bias field correction and resampling to an isotropic 1 mm³ resolution. Additionally, the dataset provides manually corrected tumor subregion segmentation from 147 patients, including enhancing tumor (ET), peritumoral edematous (ED), and necrotic tumor core (NCR).

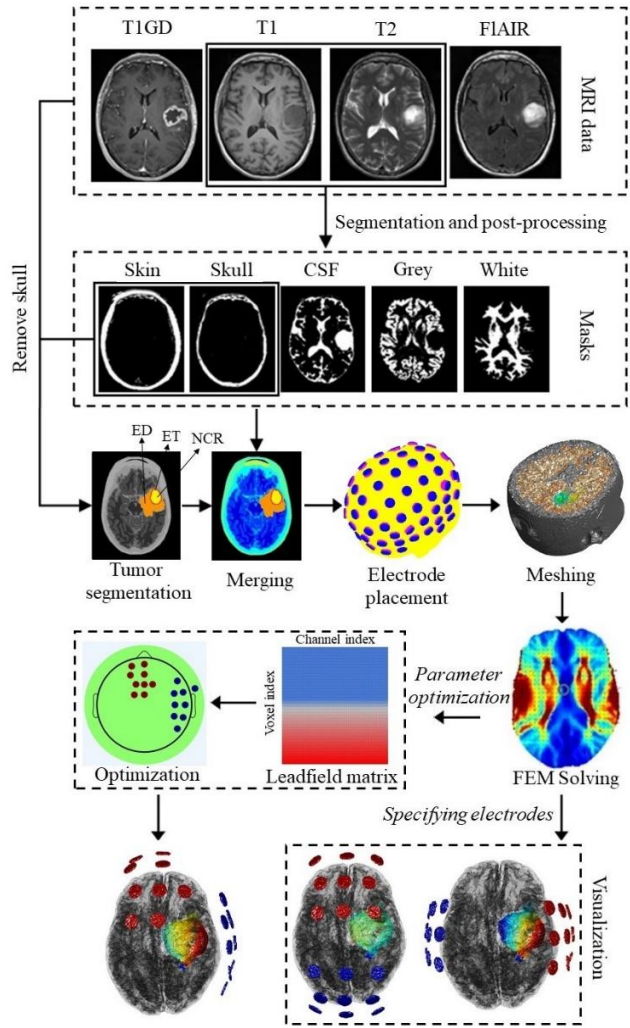

**Figure 1.** Overall pipeline of AutoSimTTF. The workflow of AutoSimTTF includes individual MRI images acquisition, normal head tissues segmentation (scalp, skull, CSF, grey matter, and white matter), tumor Segmentation (ED, ET and NCR), transducer placement, mesh generation, and finite element calculations, and parameter optimization (or specifying transducers).

*2.1.2 Normal head tissues segmentation*

In this study, T1 and T2 weighted MRI are used to segment normal head tissues, including the scalp, skull, CSF, grey matter, and white matter. The automatic segmentation tool SPM12 is used for segmentation. SPM12 utilizes a unified segmentation algorithm based on Gaussian mixture models to estimate the posterior probability of each voxel belonging to a particular tissue based on its intensity and the prior probability distribution of different tissues in MRI images. After inputting registered T1 and T2 data, selecting the desired tissue.

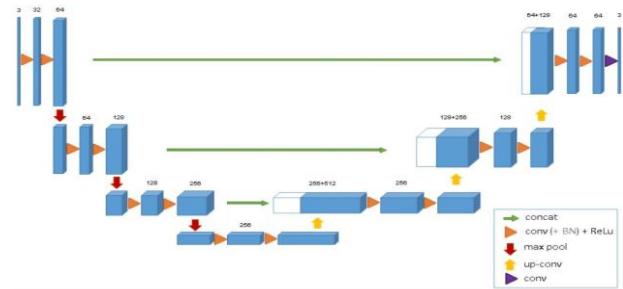

**Figure 2.** The structure of the 3D-Unet network [21]. Blue boxes represent feature maps, with their dimensions indicating the size of the feature maps.

*2.1.3 Tumor segmentation*

This study utilizes the 3D U-net for the tumor segmentation, requiring additional use of FLAIR and T1Gd images. The overall structure of the 3D U-net is U-shaped, comprising an encoding path and a decoding path, as depicted in figure 2. Feature maps in the encoding path are connected to corresponding layers in the decoding path through skip connections, which help restore precise localization information during decoding [21]. The encoding path includes multiple 3D convolutional layers and pooling layers, while the decoding path consists of up-sampling (deconvolution layers) and convolutional layers to progressively restore the spatial resolution and detailed local information of the images. This network can handle large volumes of three-dimensional medical data and capture spatial information within volumetric data, which is particularly important for brain tumors that require consideration of overall morphology and sites. Additionally, the complexity of brain necessitates global context information for better segmentation, and 3D U-net is capable of processing entire volumes rather than single slices, thereby better capturing this global information [22]. The training dataset for the 3D U-net comes from the 2021 Brain Tumor Segmentation Challenge (BraTS), which includes multi-modal MRI data from 1,251 diffuse glioma patients. Each case's data includes T1, T1Gd, T2, and T2-FLAIR modalities, with tumor subregional labels NCR, ED, and ET annotated by neuroradiology experts.

Before segmenting brain tumors, it is necessary to remove the skull to avoid signal intensity similarities between the skull and tumor in MRI, as well as to prevent segmentation from mistakenly classifying parts of the skull as the tumor due to edge similarities. By performing a difference operation between the original image and the segmentation masks for the scalp and skull, skull-removed MRI are obtained, which are then input into the 3D U-net for tumor segmentation.

### 2.1.4 Transducer placement

After segmenting the normal tissue and tumor of head, the transducer and hydrogel model needs to be constructed. MATLAB scripts are employed to determine the center coordinates of transducers and construct transducer and hydrogel models on the scalp surface. International EEG systems, such as the 10-20 system, 10-10 system, or BioSemi-256 system, could be used for determining positions of transducers. Based on predefined spatial positions in TPM (nasion (Nz), inion (Iz), preauricular points (PAR and PAL)), coordinate transformations are performed with affine matrices obtained from SPM12 segmentation of MRI, yielding actual coordinates of these reference points on the scalp surface [23]. The center coordinates of transducers are determined based on reference point and names of transducers (corresponding to positions in the transducer positioning system). The transducer and hydrogel models are then constructed on the scalp surface.

### 2.1.5 Mesh generation

Prior to finite element calculation, segmented volumetric data must be converted into a mesh model suitable for computation. This involves merging masks of segmented tissues, transducers, and hydrogel models. A series of parameters are then defined to control quality and density of mesh. Then the cgalv2m function in the iso2mesh toolkit is called for mesh subdivision, which generates mesh models based on the input images and parameters, including nodes, volumetric meshes, and surface meshes [24]. Finally, the meshes are saved in Gmsh format for subsequent finite element calculations.

### 2.1.6 Finite element calculation

After constructing a head mesh model including the tumor, brain EF distribution is calculated using the EF calculation module based on the GetDP solver. GetDP is an open-source solver that allows users to customize simulation problems by defining equations, boundary conditions, and source terms. It incorporates various solvers to handle linear and nonlinear systems, including direct and iterative solvers [25]. Before calling GetDP, it is essential to predefine parameters such as tissue properties (conductivity and relative permittivity), stimulation frequency, and current density of transducer surface.

### 2.1.7 parameter optimization

The goal of parameter optimization is to determine the transducer positions and current magnitudes for TTFields in individual patients, based on their anatomical information, so as to maximize EF intensity at the tumor site and improve field focality. Prior to parameter optimization, an EF transfer matrix (A matrix) must be calculated for all transducer positions. This matrix represents the mapping between the current magnitudes injected by the transducers and the resulting EF intensity within the brain, as illustrated in (1):

$$e = As \quad (1)$$

where $e$ represents the EF intensity at different positions, and $s$ represents the current magnitudes of different transducers. Intensity and focality of EF are optimized on the individual's head model. The optimization aims to maximize EF intensity at the target regions while minimizing it in non-target regions for better focality. To enhance computation efficiency, the minimization of EF intensity in non-target regions is represented using the following approach: selecting the farthest distance $D_m$ from the target node within the head, obtaining nodes at a quarter of this distance, and then minimizing the EF intensity at these nodes. The overall parameter optimization formula is defined as follows:

$$s = \arg\max_s (\boldsymbol{E}_{exp}s - \lambda sum(\boldsymbol{E}_{0.25Dm})) \quad (2)$$

where $\boldsymbol{E}_{exp}$ denotes the expected EF intensity distribution at the tumor sites, and $\boldsymbol{E}_{0.25Dm}$ denotes the EF intensity at positions a quarter distance from $D_m$. A weight parameter $\lambda$ balances EF intensity and focality, with higher $\lambda$ indicating better focality, and lower $\lambda$ resulting in higher EF intensity at the target region.

To ensure tolerance of stimulation and convenience of clinical application, there are several safety constraints imposing on the parameter optimization.

*a)* The sum of absolute value of currents must not exceed a specific total value $I_{total}$, shown as follows:

$$\sum |s_i| \leq I_{total} \quad (3)$$

*b)* The absolute value of the current for any single transducer must not exceed a specific value $I_{max}$, shown as follows:

$$|s_i| \leq I_{max} \quad (4)$$

*c)* The number of transducers used is constrained to $n$.

Using this method, a personalized montage for the individual patient can be obtained. Besides parameter optimization, EF simulation can also be performed by specifying the used transducers and their corresponding current magnitudes.

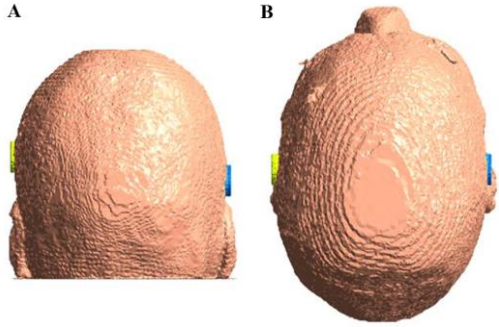

**Figure 3.** Locations of transducers: a) Rear view, b) Top view

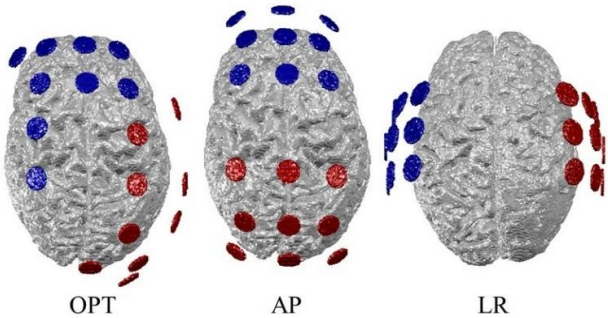

**Figure 4.** Comparison of three montages (OPT, AP, LR). OPT: results of parameter optimization by AutoSimTTF. AP: the anterior-posterior 3×3 transducer array. LR: the left-right 3×3 transducer array

TABLE I

Dielectric properties of tissues

| tissues | $\sigma$ (S/m) | | | $\varepsilon_r$ | | |
|---|---|---|---|---|---|---|
| | MIN | STD | MAX | MIN | STD | MAX |
| skin | 0.100 | 0.250 | 0.700 | 1000 | 5000 | 10000 |
| skull | 0.003 | 0.013 | 0.080 | 50 | 200 | 300 |
| CSF | 1.400 | 1.790 | 2.600 | 100 | 110 | 110 |
| grey | 0.150 | 0.250 | 0.800 | 2000 | 3000 | 3800 |
| white | 0.080 | 0.120 | 0.800 | 1000 | 2000 | 3000 |
| ED/ET | 0.200 | 0.240 | 1.600 | - | 2000 | - |
| NCR | 0.500 | 1.000 | 4.000 | - | 110 | - |
| gel | - | 0.100 | - | - | 100 | - |
| ceramic transducers | - | 2.5e-14 | - | - | 10000 | - |

## B. EFs simulation evaluation

7 MRI samples with expert manual tumor segmentation labels were randomly selected in the UPenn-GBM dataset. We performed EF calculations using both the pipeline proposed in this paper and the conventional manual segmentation process. The conventional process involved employing SPM12 for normal head tissue segmentation, merging it with the manually segmented tumor masks, and then utilizing iso2mesh to generate tetrahedral meshes from the segmented volumes. Following manual adjustments, the complete mesh model was imported into COMSOL for calculation.

Both pipeline used identical tissue dielectric properties, including conductivity σ and relative permittivity $\varepsilon_r$, as shown in Table 1. The same transducer positions were set with the anode located at T8 of the 10-20 system and the cathode at FTT7h, as depicted in figure 3. The same boundary conditions were applied: a 100 mA current was input at the anode, the cathode was grounded, and the frequency was set at 200 kHz. The experimental platform comprised a Windows 10 operating system, Intel(R) Core (TM) i7-11700 @ 2.50GHz, NVIDIA GeForce RTX 4090, and 32GB RAM; the software platforms used were PyCharm and MATLAB R2021a.

To facilitate comparison, the EF intensities calculated by COMSOL at the mesh nodes were resampled to a regular grid with the same dimensions and resolution as the original MRI. The EF intensity results were interpolated to a 240 × 240 × 155 grid with a spacing of 1 mm, which was the same as original MRI. The deviation of the EF intensity at each node was then calculated using (5):

$$E_{diff} = \frac{|E_g - E_c|}{E_c} \quad (5)$$

## C. EFs simulation evaluation

To validate the performance of the parameter optimization proposed in this paper, the optimization results (OPT) were compared with traditional TTFields montages, including the anterior-posterior 3×3 transducer array (AP) and the left-right 3×3 transducer array (LR), as shown in Figure 4. The radius of the transducers used was 8 mm, and the thickness was 1 mm. 10 MRI images from the UPenn-GBM dataset were selected for EF simulations using the three montages. In each of the montages, the current for each anode was 100 mA, and the current for each cathode was -100 mA. The optimization limits the number of transducers to be the same as the number of transducers in traditional montages (9 anodes and 9 cathodes).

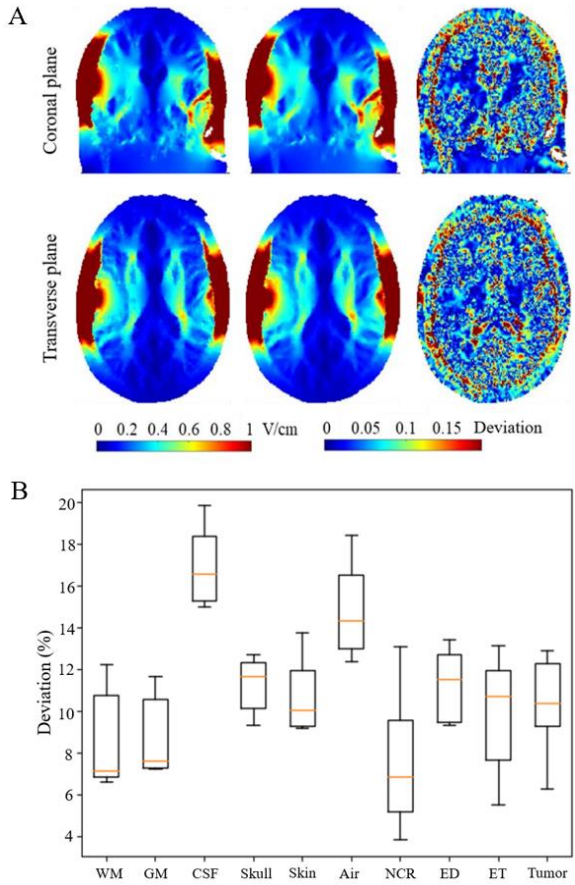

**Figure 5.** EF distribution and deviation between different simulation methods. A: the EF distribution calculated by AutoSimTTF (left) and conventional manual segmentation process (middle), and EF deviation between two methods. B: EF deviation in different tissues between two methods.

To compare the EF distributions generated by different montages in the head, two evaluation indexes were used for evaluation: *a)* The average EF intensity within the target region. *b)* The EF focality, measured by the radius $r_{0.5}$ of a sphere centred on the target point, within which the cumulative EF intensity accounts for half of the total EF intensity. A higher EF intensity and a smaller $r_{0.5}$ value indicate a better EF distribution.

## 3. Results

### 3.1 Performance of the EF simulation

The results of the EF simulations are shown in figure 5A. figure 5A (left) depicts the EF distribution calculated using the AutoSimTTF. figure 5A (middle) shows the EF distribution obtained using the control process. figure 5A (right) illustrates the deviation distribution between the two results. It can be seen that the EF distributions calculated by both methods are very similar, with deviation distributions

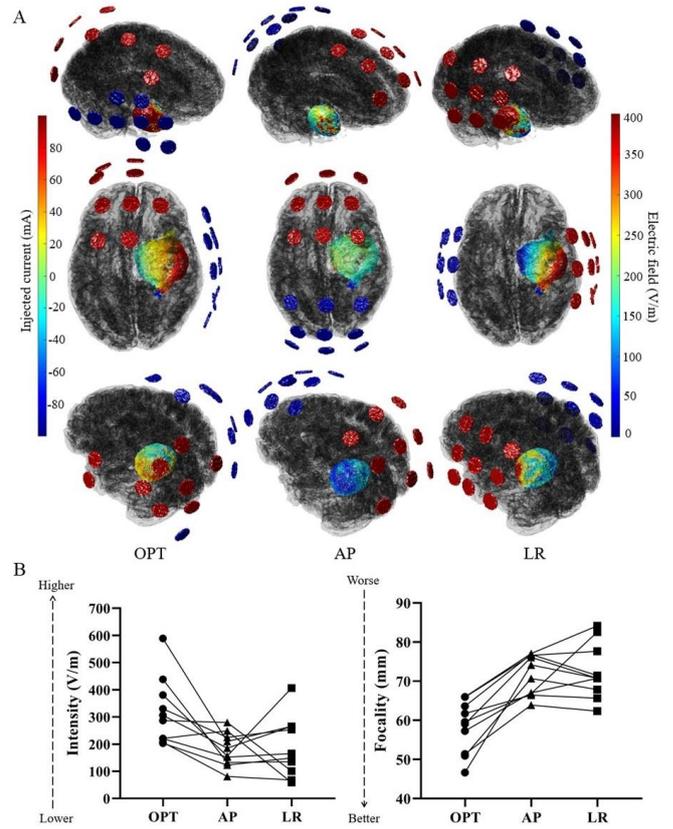

**Figure 6**. Performance of three transducer montages (OPT, AP, LR). A: EF distribution at tumor generated by three transducer montages from three patients' MRI. B: Comparison of average EF intensity at tumor (left) and EF focality (right) of three transducer montages. Higher EF intensity at tumor and lower values of EF focality indicate better EF distribution

being relatively scattered. Higher deviations are mostly located at the boundaries between CSF and other tissues.

To analyze the EF deviation in regions of different tissues, the average EF deviation for each tissue was statistically analyzed, as shown in figure 5B. The results indicate that all deviations are below 20%, with the EF deviations in most regions of tissues being below 14%. The highest deviation is found in the CSF, mainly because CSF has a high conductivity and contains many narrow regions. Before importing into COMSOL, some isolated small regions were manually discarded, and the shape of their boundaries was altered. Similarly, many isolated air regions were classified as adjacent tissues, which does not occur with the AutoSimTTF.

### 3.2. Performance of parameter optimization

Traditional montages utilize a fixed 3×3 transducer array with a preset relative positioning between the transducers, resulting in limited flexibility in transducer placement. The parameter optimization of AutoSimTTF overcomes the constraints of the fixed 3×3 transducer array, allowing each transducer to be positioned anywhere on the head without restrictions from the positions of other transducers. This approach enables the generation of more focused and effective EF distributions for tumors located in various positions, ultimately enhancing treatment outcomes.

The visualized EF distributions generated by the three montages for three patients are demonstrated in figure 6A. It is observable that the EF intensity generated at the tumor site by the optimized montages surpasses that of the two traditional montages. A comparison of the average EF intensity at the tumor sites and the EF focality for the three montages across all patients is depicted in figure 6B. The average EF intensity produced at the tumor site by the optimized montages is 92.5% higher than that of the AP montage and 111.9% higher than that of the LR montage. In terms of EF focality, the optimized montages achieved a 18.5% and 19.4% improvement over the AP and LR montages, patients with different location of GBM, the use of the optimized montages can generate a stronger and more focused EF distribution at the tumor sites in the head.

## 4. Discussion

This paper developed an automated analysis pipeline, AutoSimTTF, for TTFields, which reduces the difficulty for researchers in calculating TTFields and provides convenience to medical personnel in treatment planning of TTFields.

In the process of TTFields simulation, manual correction is required in several stages, such as MRI segmentation, transducer placement, and finite element mesh model construction. Additionally, the segmentation and modeling of the tumor region also need to be performed manually. This process is not only time-consuming but also relies on the expertise and experience of medical personnel, which affects the reliability and reproducibility of the calculations. To address these issues, The automated workflow of AutoSimTTF includes normal tissues and tumor segmentation, transducer placement, mesh generation, finite element calculation and parameter optimization (or specifying transducers). By conducting secondary development of the GetDP, the definition and calculation of the EF distribution equation in the TTFields scenario for the head were realized. The use of quasi-electrostatic equations combined with Dirichlet and Neumann boundary conditions described the problem of solving the EF for tumors, and the partial differential equations in continuous space were transformed into linear systems of equations.

This study proposes a parameter optimization specifically for TTFields, which aims to optimize two objectives: the EF intensity at the tumor site and the focality of the EF distribution. This optimization method has high computational efficiency (only a few minutes per subject). By comparing the optimized montages with traditional TTFields montages, it was observed that the optimized montages generated significantly higher average EF intensity (111.9% higher) at the tumor sites and exhibited a more focused EF distribution (19.4% better) in the head. Due to the variable positions of tumors in the head, the efficacy of traditional montages varies significantly across different patients. The proposed optimization method ensures effective treatment for tumors located at different positions. As the optimized transducer arrays are not constrained to the type of 3×3 array, designing more flexible transducer arrays that can be easily integrated with the optimization method represents a key objective for future research.

The proposed AutoSimTTF still has areas for potential research and improvement. The segmentation of tissues in the AutoSimTTF was completed in two isolated parts: normal head tissue segmentation and tumor segmentation. In the future, a unified segmentation method could be considered to improve accuracy and speed of segmentation. Additionally, given that GBM patients usually undergo TTFields after surgical tumor resection, prior information about the patient's surgical condition could be incorporated into segmentation of tissues to enhance the segmentation effect. When defining dielectric properties of tissues, this study considered tissues as homogeneous and isotropic substances. However, the actual situation is more complex. The impact of tissue heterogeneity and anisotropic dielectric properties on EF calculation results should be considered, and the actual dielectric properties of tissues should be measured as much as possible. Furthermore, this study did not use in vivo measurements to verify the effectiveness of parameter optimization results and the accuracy of computational results. Future clinical trials should demonstrate that optimized individual treatment plans improve therapeutic effects through parameter optimization, and experiments of in vivo measurement should be conducted to assess the accuracy of the computational results.

## 5. Conclusion

A fully automatic pipeline for EF simulation and treatment planning of TTFields, AutoSimTTF, is proposed in this paper. AutoSimTTF is capable of fully automatic and rapid tissue segmentation and EF simulation for individual patients with brain tumors based on their MRI. It has demonstrated its accuracy by comparing the process of traditional manual tumor segmentation. AutoSimTTF could perform parameter optimization based on the individual's head anatomical information to achieve a precise treatment plan of TTFields.

**Acknowledgements**